# The Internal and External Problems of String Theory

## -

## A Philosophical View [1] [2]

**Reiner Hedrich**[3]

Institut für Philosophie, Fakultät Humanwissenschaften und Theologie, Universität Dortmund, Emil-Figge-Strasse 50, 44227 Dortmund, Germany

Zentrum für Philosophie und Grundlagen der Wissenschaft, Justus-Liebig-Universität Giessen, Otto-Behaghel-Strasse 10 C II, 35394 Giessen, Germany

## Abstract

String theory is at the moment the only advanced approach to a unification of all interactions, including gravity. But, in spite of the more than thirty years of its existence, it does not make any empirically testable predictions, and it is completely unknown which physically interpretable principles could form the basis of string theory. At the moment, "string theory" is no theory at all, but rather a labyrinthic structure of mathematical procedures and intuitions. The only motivations for string theory consist in the mutual incompatibility of the standard model of quantum field theory and of general relativity as well as in the metaphysics of the unification program of physics, aimed at a final unified theory of all interactions, including gravity. The article gives a perspective on the problems leading to and resulting from this situation.

---

[1]     Research for this paper was supported by the Deutsche Forschungsgemeinschaft under the project FA 261/7-1 "Vereinheitlichung in der Physik durch den Superstring-Ansatz: Wissenschaftstheoretische und naturphilosophische Analyse". Preliminary research was carried out during my stay (Jan. - Apr. 2002) as a Visiting Fellow at the Center for Philosophy of Science of the University of Pittsburgh. I am grateful to both institutions as well as to Brigitte Falkenburg.

[2]     More details are to be found in Hedrich (2006b).

[3]     Email: Reiner.Hedrich@phil.uni-giessen.de  &  hedrich@fb14.uni-dortmund.de

# 1. String Theory and Philosophy of Science

Some physicists seem to take it for sure that string theory[4] is the adequate theory of quantum gravity and the ultimate theory of nature.

> *"I believe that we have found the unique mathematical structure that consistently combines quantum mechanics and general relativity. So it must almost certainly be correct."* (Schwarz (1998) 2)

Others see in it the ultimate hype in which the ideals of physics as an empirical science get lost.

> *"The theory has been spectacularly successful on one front, that of public relations."* (Woit (2001) 1)[5]

But what is string theory really? - There are certainly good reasons to claim that this question can not be answered conclusively at the moment, because string theory isn't yet a complete theory.

> *"Elegance requires that the number of defining equations be small. Five is better than ten, and one is better than five. On this score, one might facetiously say that String Theory is the ultimate epitome of elegance. With all the years that String Theory has been studied, no one has ever found even a single defining equation! The number at present count is zero. We know neither what the fundamental equations of the theory are nor even if it has any. Well then, what is the theory, if not a collection of defining equations? We really don't know."* (Susskind (2005) 124)

Already in 1988, after the first wave of string enthusiasm which started with the development of the first consistent anomaly-free perturbative supersymmetric string theories, Robert Weingard wrote in his paper *A Philosopher Looks at String Theory*:

> *"[...] there is, in a sense, no theory for the philosopher to analyse."* (Weingard (1989) 138)

In a certain way, the conditions for an evaluation of string theory did not change a lot since Weingard's assessment - an assessment which did not prevent him from writing his paper on the subject. Although string theory is at present the most extensively developed approach to a future theory of quantum gravity, it seems to be a rather confusing collection of physical intuitions and mathematical procedures which either will or will not lead finally to a physical theory. At the moment, string theory has no clear and unambiguous nomological basis; no physically motivated fundamental principle is known for string theory.

> *"Ironically, although superstring theory is supposed to provide a unified field theory of the Universe, the theory itself often seems like a confused jumble of folklore, random rules of thumb, and intuitions. This is because the development of superstring theory has been unlike that of any other theory [...]. Superstring theory [...] has been*

---

[4]    A systematical introduction to string theory can be found in Polchinski (2000), (2000a), Kaku (1999), Lüst / Theisen (1989) and Green / Schwarz / Witten (1987). For more recent developments, see especially Lerche (2000), Schwarz (2000), Dienes (1997) and Vafa (1997). The early development of string theory is reflected in a commented collection of original articles: Schwarz (1985). Greene (1999) gives a recommendable popular introduction.

[5]    See also Woit (2002) and Schroer (2006).

*evolving backward for the past 30 years. It has a bizarre history [...]. [...] physicists have ever since been trying*
*to work backward to fathom the physical principles and symmetries that underlie the theory. [...] the*
*fundamental physical and geometrical principles that lie at the foundation of superstring theory are still*
*unknown."* (Kaku (1999) vii f)

And moreover, on the one hand, string theory does not make any predictions which could be tested empirically.[6]

*"The experimental situation is best described with Pauli's phrase 'it's not even wrong'. [...] String theory not only*
*makes no predictions about physical phenomena at experimentally accessible energies, it makes no predictions*
*whatsoever."* (Woit (2001) 2)

On the other hand, there are no empirical data which would make a theory of quantum gravity necessary; not the least empirical fact is known which would not be completely compatible with general relativity or the standard model of quantum field theory.

*"We have today a group of fundamental laws, the standard model and [general relativity], [...] there aren't today*
*experimental facts that openly challenge or escape this set of fundamental laws."* (Rovelli (1998) 2)

All motivations for a theory of quantum gravity are based on conceptual considerations which come from the traditionally successful program of unification,[7] either in its merely conceptual (model-theoretical) or in its full-blown nomological form.

*"[...] with absolutely no experimental basis, string theorists constructed a monumental mathematical edifice."*
(Susskind (2005) 270)

Without any empirical motivation, confirmation or control, it is completely unclear, if the rather unquestionable conceptual success of string theory should be seen as an indication that it will lead in the future to an adequate description of nature. The conceptual success could as well be an artifact of a specific methodology (based on specific mathematical models and procedures) which finally will lead to a dead end. - So, wouldn't it mean an incalculable as well as unnecessary risk for a philosopher of science to deal with such an unfinished research program which has not reached (yet) any contact to empirical data?

*"No data, no theory, no philosophy?"* (Butterfield / Isham (2001) 36)

Postponing a philosophical view on string theory and its implications and problems until the consolidation of the research program (in form of a physically motivated, nomologically unambiguous and empirically controlled theory) would - one could think - lead to more adequate conditions for an evaluation. Or any further examination would turn out to be obsolete, because the research program would already have vanished.[8] - So, why should a philosopher of science be interested in string theory at the present? Why shouldn't we simply wait for the complete theory?[9]

---

[6]     Cf. Hedrich (2002) and (2002a).
[7]     Cf. Weinberg (1992).

String theory exists now for more than thirty years. For more than twenty years it is now an aspiring research program taken very seriously by more and more physicists. Meanwhile a tremendous number of researchers is working on string theory. This is all the more astonishing since string theory did not make the least empirically testable prediction during all the decades of its existence. - How can an empirical-scientific research program (string theorists understand themselves as physicists) survive for such a long time without the least empirical confirmation? Could it be that string theory, even if it was created by physicists, is not really a part of physics or of the empirical sciences? Has physics with its unification program and under strict pursuance of its traditional methodological strategies possibly transcended the context of the empirical sciences and entered that of a mathematically inspired metaphysics of nature?

An adequate answer to these questions is certainly not to be expected from physics; physics does not have the methodology to answer such metatheoretical questions. These are undoubtedly questions to be answered in the context of philosophy of science. And there are no good reasons to wait. There are no good reasons to put off a necessary metatheoretical investigation of string theory for some decades, in the hope that then there will be a complete, empirically vindicated theory or that the problem sorts itself out on its own. The problem will not necessarily sort itself out on its own. The reasons for that are to be found in the methodological and strategical foundations of physics. And those reasons will only be comprehensible in the context of a metatheoretical investigation. Philosophy of science should not lose the contact to the actual problems of physics. It is necessary to understand the motivations and developments which led to string theory, its actual state, its structure, its problems, as well as its prospects for the future.[10]

---

[8]     Anyway, if the ambitions of string theory will turn out to be successful or not, the theory and its development will be of eminent interest for historians of science, and especially for sociologists of science.

[9]     There are probably many more good reasons for a philosophical examination of string theory than those mentioned in the following.

[10]    But a philosophical investigation of incomplete theories like string theory or even the complete field of approaches to quantum gravity has to be much more flexible and open-minded than the traditional methodologies of philosophy of science might suggest:

*"'Quantum gravity' primarily refers to an area of research, rather than a particular theory of quantum gravity. Several approaches exist, none of them entirely successful to date. Thus the philosopher's task, if indeed she has one, is different from what it is when dealing with a more-or-less settled body of theory such as classical Newtonian mechanics, general relativity, or quantum mechanics. In such cases, one typically proceeds by assuming the validity of the theory or theoretical framework and drawing the ontological and perhaps epistemological consequences of the theory, trying to understand what it is that the theory is telling us about the nature of space, time, matter, causation, and so on. Theories of quantum gravity, on the other hand, are bedeviled by a host of technical and conceptual problems, questions, and issues which make them unsuited to this approach. However, philosophers who have a taste for a broader and more open-ended form of inquiry will find much to think about."* (Weinstein (2005) 2)

## 2. The Internal and the External Problems of String Theory

String theory is certainly not a physical theory. But should it at least be understood as a physically motivated model-theoretical approach on its way to become a physical theory in the future? Or is it only a mathematical game claiming to be or to become a physical theory without any justification? - The assessments to be found on this matter in two of the best-known textbooks of string theory are astonishingly clear-minded and unpretentious. Obviously even some of the acknowledged experts see string theory primarily as a formal mathematical construct, although motivated by physical thinking:

> *"[...] the developments we have presented in string theory have been purely formal."* (Kaku (1999) 337)

> *"String theory is a rich structure, whose full form is not yet understood. It is a mathematical structure, but deeply grounded in physics. It incorporates and unifies the central principles of physics: quantum mechanics, gauge symmetry, and general relativity, as well as anticipated new principles: supersymmetry, grand unification, and Kaluza-Klein theory. Undoubtedly there are many remarkable discoveries still to be made."* (Polchinski (2000a) 429)

Probably the best way to understand what string theory is, is to ask: What are its motivations, how did it develop, and what are its problems? - Some of the answers to these questions might appear astonishing. String theory is *not* the result of an intended and planned theory development, starting from clear physical problems or even empirical data which could not be reproduced by the established theories. It is *not* a mathematical formulation or implementation of basic physical intuitions or principles. Rather it is the result of a sequence of sometimes bizarre coincidences, and of mathematical strategies and procedures, introduced to deal with these coincidences.

The central discovery of string theory is that the quantization of the dynamics of a relativistic string leads to spin-2 states, besides many other states. It was this discovery and the interpretation of the spin-2 states as gravitons which determined the actual and only physical relevance of string theory. Before the discovery of spin-2 states, string theory was part of hadron physics. It was the casual discovery of a correspondence between hadron scattering behaviour and Euler's beta-function which led Gabriele Veneziano in 1968 to a model which could be identified in 1970 by Leonard Susskind, Holger Nielsen, Yoichiro Nambu and T. Goto as describing the quantized dynamics of a relativistic string. But this 26-dimensional bosonic string theory was not very successful in the context of hadron physics and lost its relevance after the rise of the quark model. The discovery of gravitons as string states, made by Joel Scherk and John Schwarz in 1974, led - with a shift in the energy scale of about 15 powers of ten[11] - to an ad-hoc relocalization of string theory into the context of quantum gravity and that of the program of a nomological unification. It mutated from an unsuccessful theory in hadron physics to a prospective unified theory of all interactions, including gravitation as well as interactions mediated by spin-1 gauge fields. And finally, in the ten-dimensional supersymmetric formulation of string theory it also included fermionic matter fields.

---

[11]   String theory reproduces general relativity as well as gauge invariances, possibly those of the standard model of quantum field theory, as low-energy approximations; general relativity comes with the phenomenologically correct parameter values, if the string length and tension are assumed to lie in the order of magnitude of the Planck scale: a further indication for the context of quantum gravity.

Exclusively the fact that the quantized string has spin-2 states which can be interpreted as gravitons and which are completely unmotivated in the context of hadron physics led to this change of strategy and of context from hadron physics to quantum gravity. String theory is not the result of an intended approach to a unified description of all interactions, but fell into this role by a casual discovery. It is not the result of a planned development of a theory aimed at the description of a certain context of relevance or at the solution of specific physical problems, but rather that of a casual finding of a theory: a mathematical construct which, after having lost its original physical relevance, found by chance a new context of relevance, without even looking for it.

In its new role as a realization of an all-encompassing nomological unification, all the motivations for the unification program could be claimed for string theory. These were, on the one hand, the problems of the conceptual incompatibility of our established, most fundamental theories of physics, i.e. general relativity and quantum (field) theory. On the other hand, the former success of the unification of electromagnetic and weak interactions (the *Glashow-Salam-Weinberg model*) reinforced the hope for a solution of the still virulent problems with a quantization of gravity in the broader context of a nomologically unified theory of all interactions; all former attempts at a quantization of gravitation had led inevitably to non-renormalizability. String theory thereby became the new hope for the next step in the theoretical implementation of an old philosophical intuition, an intuition which had, at least since Newton, already seen a long sequence of successes in physics: the idea of the unity of nature.

And string theory had no other option. Its only possible justification of existence lies in the unification program of physics. The existence of gravitons as well as spin-1 gauge bosons is only a promising feature, if string theory plays the role of an all-encompassing, nomologically unified theory.

But did string theory play this role successfully after all? - There are some doubts with regard to that. What has string theory finally to offer except for a unified, but perturbative mathematical scheme including spin-2 states, interpretable as gravitons, as well as gauge boson, matter, and scalar fields? That string theory reproduces formally general relativity and the corresponding phenomenology of gravitation is - as even string theorists meanwhile know - not much of a surprise; it is a direct implication of the existence of spin-2 states.[12] Apart from that, string theory consists of

---

[12]        At the time of their discovery it wasn't completely clear for string theorists that the dynamics of spin-2 states necessarily lead to the reproduction of general relativity and of the phenomenology of gravitation as a low-energy approximation:

*"[...] with appropriate caveats, general relativity is necessarily recovered as the low-energy-limit of* any *interacting theory of massless spin-2 particles propagating on a Minkowski background, in which the energy and momentum are conserved (Boulware and Deser 1975)."* (Butterfield / Isham (2001) 59)

*"[...] a general result that goes back to Feynman: any theory of an interacting spin two massless particle must describe* gravity. *So string theory must reproduce gravitational physics."* (Giddings (2005) 6)

*"Weinberg (1995), in his discussion of covariant quantum gravity, showed that, in the vacuum case, one can* derive *the equivalence principle and general relativity from the Lorentz-invariance of the spin-2 quantum field theory of the graviton: the spin-2 theory is* equivalent *to general relativity and* follows *from the quantum theory.*

a sometimes astonishing and sometimes mathematically highly interesting labyrinth of model-theoretical constructs in search for a physically motivated principle which could justify the approach and give it a fundament. And it is completely dominated by *internal problems*: conceptual, model-theoretical and mathematical problems which came only into existence with the theory under development - problems which did not exist before.

*External problems* - in the sense of original physical problems which (i) serve as an independent motivation for the development of a theory or (ii) which make a theory at least interesting, because it promises their solution - are completely in the backdrop for string theory. There are especially no external problems in the sense that they triggered the development of string theory as a unified theory. And there is only one decisive external problem that gives at least a post-hoc motivation for string theory: it is the fact that string theory led by chance to the discovery of the graviton and promised thereby a solution to the external problem of finding a theory which consistently unifies general relativity and quantum mechanics.[13]

With regard to all other possibly relevant external physical problems - problems for which a theory of quantum gravity should be able to find a solution - there is none. String theory does not lead to any deeper insights with regard to the nature of space and time. It does not lead to an understanding of the quantum features of gravitation. It does not explain the equivalence of gravitational and inertial mass, or give new answers to the question what mass is.[14] - It is scarcely a far-fetched idea to expect solutions to these problems from a theory of quantum gravity.

String theory - which promises to give an all-encompassing, nomologically unified description of all interactions - did not even lead to any unambiguous solutions to the multitude of explanative desiderata of the standard model of quantum field theory: the determination of its specific gauge invariances, broken symmetries and particle generations[15] as well as its 20 or more free parameters, the chirality of matter particles, etc.

---

*The upshot of this is that any theory with gravitons is a theory that can accommodate general relativity (in some appropriate limit). This analysis forms the basis of string theory's claim that it is a candidate theory of quantum gravity: since there is a string vibration mode corresponding to a massless spin-2 particle, there is an account of general relativity [...]."* (Rickles (2005) 9)

[13]    Additionally, also by pure chance, it promises to give a nomologically unified description of all interactions. But without the least experimental results which would make a unified theory necessary, the only concrete motivation for a unification (beside metaphysical considerations) consists in the mutual incompatibility of general relativity and quantum theory; and the desire to overcome this incompatibility is sufficient at best as a motivation for a conceptual unification: a theory of quantum gravity or quantum geometry like e.g. *Loop Quantum Gravity* (see below). It does not make necessary a nomological unification of all interactions in the sense of string theory.

[14]    The equivalence of inertial and gravitational mass is one of the fundaments of general relativity. General relativity presupposes this empirical fact in the determination of its theory structure. It does not explain it in any way. But a fundamental theory should explain why the inertial mass (a quantity of motion) is equivalent to the gravitational mass (the "charge" of gravitation).

[15]    String theory does at least give an explanation for the existence and for the number of particle generations. The latter is determined by the topology of the compactified additional spatial dimensions of string theory; their topology determines the structure of the possible oscillation spectra. The number of particle generations is identical to half the absolute value of the Euler number of the compact Calabi-Yau topology. But, because it is completely unclear which topology should be assumed for the compact space, there are no definitive results. This ambiguity is part of the *vacuum selection problem*; there are probably more than $10^{100}$ alternative scenarios in the so-called *string landscape* (see

*"Sal - [...] There are other open problems of the standard model [...]. Like understanding why there are three families. Does string theory solve that?*
Simp - ... no ...
Sal - *why the cosmological constant is small ?*
Simp - ... no ...
Sal - *giving a better account of symmetry breaking ?*
Simp - ... no ... ."* (Rovelli (2003) 4)

Attempts at a concrete solution of the relevant external problems (and explanative desiderata) either did not take place, or they did not show any results, or they led to escalating ambiguities and finally got drowned completely in the *string landscape* scenario: the recently developed insight that string theory obviously does not lead to a unique description of nature, but describes an immense number of nomologically, physically and phenomenologically different worlds - with different symmetries, parameter values, values of the cosmological constant, etc.[16]

String theory seems to be by far too much preoccupied with its internal conceptual and mathematical problems to be able to find concrete solutions to the relevant external physical problems. It is almost completely dominated by internal consistency constraints.

It is *not* the fact that we are living in a ten-dimensional world which forces string theory to a ten-dimensional description. It is that perturbative string theories are only anomaly-free in ten dimensions; and they contain gravitons only in a ten-dimensional formulation. The resulting question, how the four-dimensional spacetime of phenomenology comes off from ten-dimensional perturbative string theories (or its eleven-dimensional non-perturbative extension: the mysterious, not yet existing *M theory*), led to the compactification idea and to the braneworld scenarios - and from there to further internal problems.

It is *not* the fact that empirical indications for supersymmetry were found, that forces consistent string theories to include supersymmetry. Without supersymmetry, string theory has no fermions and no chirality, but there are tachyons which make the vacuum instable; and supersymmetry has certain conceptual advantages: it leads very probably to the finiteness of the perturbation series, thereby avoiding the problem of non-renormalizability which haunted all former attempts at a quantization of gravity; and there is a close relation between supersymmetry and Poincaré invariance which seems reasonable for quantum gravity. But it is clear that not all conceptual advantages are necessarily part of nature - as the example of the elegant, but unsuccessful *Grand Unified Theories* demonstrates.

Apart from its ten (or eleven) dimensions and the inclusion of supersymmetry - both have more or less the character of only conceptually, but not empirically motivated ad-hoc assumptions - string theory consists of a rather careful adaptation of the mathematical and model-theoretical apparatus of

---

below). Moreover all concrete models, deliberately chosen and analyzed, lead to generation numbers much too big. (There are phenomenological indications that the number of particle generations can not exceed three. String theory admits generation numbers between three and 480.)

[16]     Cf. Hedrich (2006) as well as (2005), (2005a) and (2006a). Cf. also Banks / Dine / Gorbatov (2003), Dine (2004), Douglas (2003), Susskind (2003), (2004) and (2005).

perturbative quantum field theory to the quantized, one-dimensionally extended, oscillating string (and, finally, of a minimal extension of its methods into the non-perturbative regime for which the declarations of intent exceed by far the conceptual successes). Without any empirical data transcending the context of our established theories, there remains for string theory only the minimal conceptual integration of basic parts of the phenomenology already reproduced by these established theories. And a significant component of this phenomenology, namely the phenomenology of gravitation, was already used up in the selection of string theory as an interesting approach to quantum gravity. Only, because string theory - containing gravitons as string states - reproduces in a certain way the phenomenology of gravitation, it is taken seriously.

But consistency requirements, the minimal inclusion of basic phenomenological constraints, and the careful extension of the model-theoretical basis of quantum field theory are not sufficient to establish an adequate theory of quantum gravity. Shouldn't the landscape scenario of string theory be understood as a clear indication, not only of fundamental problems with the reproduction of the gauge invariances of the standard model of quantum field theory (and the corresponding phenomenology), but of much more severe conceptual problems? Almost all attempts at a solution of the internal and external problems of string theory seem to end in the ambiguity and contingency of the multitude of scenarios of the string landscape.[17] That no physically motivated basic principle is known for string theory and its model-theoretical procedures might be seen as a problem which possibly could be overcome in future developments. But, what about the use of a static background spacetime in string theory which falls short of the fundamental insights of general relativity and which therefore seems to be completely unacceptable for a theory of quantum gravity?[18]

At least since the change of context (and strategy) from hadron physics to quantum gravity, the development of string theory was dominated by internal problems which led with their attempted solutions to further internal problems, and so on. The result of this successively increasing self-referentiality is a more and more enhanced decoupling from phenomenological boundary conditions and necessities. The contact with the empirical does not increase, but gets weaker and weaker. The result of this process is a labyrinthic mathematical structure with a completely unclear physical relevance. Often it is not even obvious, if certain components of string theory are to be seen as part of the same picture, or as part of complementary formulations, or as part of competitive alternatives or scenarios.

And there are the first symptoms of a self-immunization of string theory against (the absence of) empirical control. The modalities of this self-immunization can be seen at work most concretely with regard to the search for empirical indications for supersymmetry. What will happen, if no such indications will be found in the experiments with the new accelerator generation, e.g. the Large Hadron Collider at CERN? What will string theorists do?

> *"So we are inclined to call* supersymmetry *a generic prediction. Suppose that the LHC rules this out. Will we still believe in this approach? I can only speak for myself, though I suspect that most others working in this field would agree. I believe that we have found the unique mathematical structure that consistently combines quantum mechanics and general relativity. So it must almost certainly be correct. For this reason, even though I do expect*

---

[17]     Cf. Hedrich (2006).

[18]     Cf. Smolin (2001), (2003), (2005), Rovelli (1998), (2004) and Ashtekar (2005).

*supersymmetry to be found, I would not abandon this theory if supersymmetry turns out to be absent."* (Schwarz (1998) 2)

The strategy is rather simple. String theory does not make any predictions for the masses of supersymmetric particles. Should there be no indications for these particles, one could simply insist that, obviously, they have masses beyond the range of the experimental device.

> *"[...] even if superpartner particles are not found by the Large Hadron Collider, this fact alone will not rule out string theory, since it might be that the superpartners are so heavy that they are beyond the reach of this machine as well."* (Greene (1999) 222)

Such an argumentation is certainly in no way acceptable for a physical theory. Most ironical in this case is the fact that a shortcoming of the theory, the inability to make quantitative predictions, is used as a strategical advantage for its self-immunization. This is probably something new in physics.

# 3. A Mathematically Inspired Metaphysics of Nature ?

As a mathematical construct without any empirical control, string theory seems to transcend by means of its increasing self-immunization more and more the context of physics, to become finally a form of mathematically inspired metaphysics of nature. There are many indications of a return to the modalities of an ancient or even a presocratic thinking about nature, although achieved indirectly through a highly complex mathematical apparatus.

Here it should be remembered that the ancient philosophy of nature with its rejection of any empirical control can be characterized by two fundamental assumptions: (i) that our rationality enables us to understand nature by means of pure reflection, without any necessity of empirical research, and (ii) that experiments do not make accessible nature (*physis*) as it is, but, quite the opposite, they produce artifacts which cannot be identified with nature. The first steps extending beyond this ancient conception of *physis* were developed during the era of hellenism, especially by Archimedes whose *mechanike techne* was aimed at an inventive outsmarting of nature. It was based on the idea of an empirical access to nature; and it can be seen as a predecessor to the modern empirical sciences, developed by Galileo Galilei and his contemporaries. Since Galileo, the empirical controllability of our theoretical endeavours was one of the basic requirements for science and especially for physics.

Although it is carried out with an immense mathematical extravagance, the return to a metaphysics of nature, becoming manifest now in string theory, has to be seen as a retrograde step behind the innovations which were developed and practiced already by Archimedes and which became in the 17th century the methodological basis of the empirically as well as mathematically oriented modern sciences. String theory can be understood as an almost complete revival of the ancient ideal of an exclusive penetration of nature by means of our intellect (admittedly a mathematically highly instructed intellect), without any help of observations or experimental devices, only led by means of a minimal heuristic inclusion of empirical data from the now probably ending heyday of an empirically guided access to nature.

To see only a massive empirical underdetermination in the results of this return to metaphysics would clearly be an understatement. Dominated by mathematical and model-theoretical predispositions which lead to labyrinthic ramifications, and dominated as well by metaphysical motivations and prejudices, string theory incorporates a significant degree of arbitrariness with regard to its claim to be a physical theory. Only empirical guidance could change this situation.

Jeremy Butterfield and Christopher Isham emphasize that this highly problematic situation is the case for the complete context of quantum gravity. According to their assessment, the immense self-referentiality to be found in theory construction in this field of research is foremost a consequence (i) of the non-existence of empirical data which could serve as an instrument of control for the development of theories, (ii) of the significance of metaphysical assumptions and predispositions, as well as (iii) of the mathematical and model-theoretical apparatus which the respective theories chose to apply at first and never changed significantly.

> *"[...] theory-construction inevitably becomes much more strongly influenced by broad theoretical considerations, than in mainstream areas of physics. More precisely, it tends to be based on various* prima facie *views about what the theory* should *look like - these being grounded partly on the philosophical prejudices of the researcher concerned, and partly on the existence of mathematical techniques that have been successful in what are deemed (perhaps erroneously) to be closely related areas of theoretical physics, such as non-abelian gauge theories. In such circumstances, the goal of a research programme tends towards the construction of abstract theoretical schemes that are compatible with some preconceived conceptual framework, and are internally consistent in a mathematical sense. / This situation does not just result in an extreme 'underdetermination of theory by data', in which many theories or schemes, not just a unique one, are presented for philosophical assessment. More problematically, it tends to produce schemes based on a wide range of philosophical motivations, which (since they are rarely articulated) might be presumed to be unconscious projections of the chthonic psyche of the individual researcher - and might be dismissed as such!"* (Butterfield / Isham (2001) 38)

The self-referentiality of string theory is obviously far advanced. Admittedly, one can not say that the metaphysical predispositions which became relevant as motivational background with its change of context and strategy from hadron physics to the unification program are very idiosyncratic, nor that they are tacit assumptions. But the mathematical basis of string theory - essentially an extended version of the apparatus of quantum field theory - did not change significantly during the evolution of string theory. The development of the theory was always determined by the viability of its procedures under the constraints of the once chosen mathematical basis. Indeed, such a theoretical approach probably tells us more about the theoretician's prejudices than about nature, at least as long as no empirical control can be established - or as soon as the necessity of an empirical control loses more and more its significance as a fundamental requirement, e.g. replaced by the requirement of internal consistency. In such a situation the even more problematic and ambiguous concepts of the "beauty" or the "elegance" of a theory do not help at all, even less than the consistency requirement or the metaphysical prejudices underlying the development (or selection) of the theory. Exclusively conceptual boundary conditions are not sufficient as a motivation for a physical theory. One does not reach nature in this way. However fascinating, beautiful and elegant a theory may be, without any empirical control, nature could have chosen otherwise, as Carlo Rovelli reminds us:

> *"Contrary to what sometimes claimed, the theoretical developments that have followed the standard model, such as for example supersymmetry, are* only *fascinating but non-confirmed* hypotheses. *As far we really know, nature may very well have chosen otherwise."* (Rovelli (1998) 5)

# 4. The Metaphysics of Unification

String theorists often refer to the *Glashow-Salam-Weinberg* (GSW) *model* of electroweak interactions as a motivation for string theory and as an argument for the descriptive or strategic adequacy of the unification program and its concrete realizations. The GSW model is doubtless one of the best examples of a successful nomological unification. However, one has to remember that the GSW model was not successful because it consisted in a nomological unification of electromagnetic and weak interactions or because it was the first renormalizable theory to describe weak interactions, but because it made empirically testable numerical predictions which were soon to be confirmed by experiments at CERN. No-one, not even its inventors, took the GSW model seriously because it was a unified theory, but because of its empirical success.

Without any empirical corroboration, string theory can hardly be compared to (or even seen on the same level with) the empirically successful nomological unifications like the GSW model. Why should anybody take string theory seriously? In contrast to the GSW model, it does not make the least quantitative predictions which could be subjected to an independent empirical test. The program of a nomological unification and its success in the case of the GSW model can scarcely be seen as a justification for the assumption that string theory describes nature, as long as its problems with empirical control do not change.

> *"[...] unity should not be linked to truth or increased likelihood of truth; unification cannot function as an inference ticket."* (Morrison (2000) 232)

In contrast to all former hopes, the constraints resulting from the unification program are not even strong enough in string theory to lead to an unambiguous theory structure, not even if one clings to a specific mathematical apparatus for its implementation.[19]

> *"At least up till now, the hope that unification would lead to a unique theory has failed, and it has failed dramatically. So it seems unlikely that the problem of accounting for the values of the parameters of the standard models of particle physics and cosmology will be solved by restrictions coming from the consistency of a unified theory."* (Smolin (2004) 11)[20]

---

[19]     Cf. Hedrich (2006).

[20]     As Lee Smolin tells us, this is actually no great surprise:
*"But is unification enough of a criteria to pick out the right theory? By itself it cannot be, for there are an infinite number of symmetry algebras which have the observed symmetries as a subalgebra."* (Smolin (2005) 26)

But, is the motivation given by the unification program at least sufficient to take string theory seriously, as a step on the road to a physically adequate theory of nature? Or could it be that the idea of an all-encompassing nomological unification and the motivations resulting from it for the development (or casual finding) of theories are not really an advantage in the context of quantum gravity?

Already in the late eighties, when string theory became popular, Robert Weingard contradicted the view that the unification idea on its own should be seen as a sufficient basis for a successful development of a physical theory. He pointed out to some of the most prominent failures of the unification program.

> *"But one thing we have learned from the many attempts at unified theories in this century, I would argue, is that unification, in itself, is not a guide to a successful theory. / Consider the attempts by Einstein, Schrödinger, Misner and Wheeler, and others to fashion a unified geometrical theory of gravity and electromagnetism. All of these were essentially mathematical investigations, [...] there were no empirical or theoretical clues as to how such a unification should take place."* (Weingard (1989) 150)

Einstein's futile attempt at a formulation of a *Unified Field Theory* is certainly one of the most prominent examples of such an unsuccessful offspring of the unification program. But one should also remember the *Grand Unified Theories*, developed to achieve a nomological unification of electroweak and strong interactions; this step beyond the only conceptually, but not nomologically unified standard model led to contradictions between numerical predictions for proton decay and the relevant empirical data; and it ended with a great disillusionment in high energy physics.

> *"[...] history is full of beautiful hypotheses later contradicted by Nature."* (Rovelli (1998) 2)

What are the reasons for these failures of the unification program? Could it be that the idea of an ontic unity of nature - the core concept behind the unification program - is simply wrong? Then the apparent successes of the unification program would have to be understood as the result of a close connection between the respective theories and the empirical data, and not as a consequence of the adequacy of the metaphysical idea of an ontic unity of nature!

After all, there does not even exist a consensus with regard to the question in what this unity of nature consists. And it is even less clear how and to what extent such a unity should express itself within our theoretical descriptions of nature, reflecting this ontic unity epistemically. Not even a complete ontic unity of nature would necessarily lead to an all-encompassing and adequate description of nature. There might be limits to our epistemic endeavours which could make impossible any complete exploitation of an ontic unity for our scientific intentions. Those limits could even make impossible any complete capture of nature.[21] Taking this into account, it becomes clear that the idea of an ontic unity will always remain a metaphysical hypothesis.

Consequently, in philosophy of science, there can be found a widespread scepticism (i) with regard to the philosophical motivations behind the idea of a unity of nature as well as (ii) with regard to its fertility for science and its descriptive aims. Nancy Cartwright[22] is probably the most prominent of those who doubt that nature can be adequately described by means of fundamental and nomologically unified physical theories. According to her view, reality is something which can only be captured approximately, by a patchwork of effective theories which have only a limited reliability for a specific context. These effective theories will not even have to be completely compatible to each other. So, even the requirement of a conceptual, model-theoretical uniformity and consistency might be to strong.

---

[21]     Cf. Hedrich (1990), (1995), (1998), (1998a), (1999), (2001) and (2002b).
[22]     Cf. Cartwright (1994), (1999) as well as (1983) and (1989).

Let's suppose we had a consistent theory of quantum gravity which is, at the same time, a nomologically unified description of all interactions. Let's suppose furthermore that we could rule out definitively the possibility of other theories which would reproduce general relativity and quantum theory, and the phenomenology already reproduced by these well established predecessors. So, there would not be any possibility of a mere conceptual unification without a nomological unification. Our theory would be the only consistent extrapolation from its established predecessors and their empirical content.

But, let's assume that this unique consistent extrapolation into the realm of quantum gravity does not have any independent empirical content. It does not make any predictions differing from its predecessors or transcending them. Consequently, there are no independent empirical tests for this theory.

Such a theory had to be seen as the unique mathematical construct surmounting the conceptual incompatibilities of its predecessors. It would be the only possible consistent structural extrapolation from these predecessors into the realm of quantum gravity. But, as long as there are no independent empirical corroborations, this theory could be interpreted realistically with regard to its postulated entities or to its propositions. It would not necessarily describe our world. It would not necessarily have any descriptive power. Even a conceptual extrapolation of established theories proven to be unique does not necessarily have a correlate in nature. The idea of a unity of nature could simply be wrong for our world. Reality could well turn out to be a patchwork of rather independent phenomenological areas, at least when we try to describe it by means of our epistemic capacities. It could be something which can not be described with empirical adequacy by means of coherent, unified, fundamental physical theories, but rather by a collection of effective theories which would find their relevance in a direct and close coupling to specific phenomenal areas. In this case, even if we would be able to unify by force, the results of this unification would not describe reality, they would be mere conceptual extrapolations without descriptive content.

Although there exists a long sequence of successes in the past, from Newton to the GSW model, the unification program and its idea of an ontic unity of nature could well turn out to be finally inadequate and even disadvantageous for the future prospects of physics. Metaphysical motivations alone do not lead necessarily to any descriptive successes in the empirical sciences. Postulating the existence of a real correlate of the results of a conceptually successful theoretical unification would only be justified, if there were independent empirical vindications for the unified theory. But even then the patchwork scenario could never be ruled out definitively, because a nomologically unified theory will, like every empirical-scientific theory, always be inductively underdetermined, no matter how many empirical confirmations one could find. But with empirical confirmations for the unified picture, the patchwork scenario would at least seem less natural in comparison.

# 5. The Future of String Theory  -  The Future of Physics ?

String theory has repeatedly cut the ground from under the feet of its critics by its conceptual metamorphoses. But, finally, the alternatives seem to be determined unambiguously: String theory will either be successful as a nomologically unified theory of all interactions or it will be completely irrelevant as a description of nature.

> *"Either this theory turns out to be a dead end without any relation to the physical world [...] or it constitutes a final theory in the sense that it describes all possible experimental data based on a set of foundational principles without any adjustable fundamental parameters."* (Dawid (2003) 21)

But these alternatives cover only the possible outcome of the pending proceedings. But what, if these pending proceedings do not lead to any outcome? It could well be that string theory is already in a situation in which it can not really fail in any of the ways in which former attempts at a nomological unification like the geometrisation program or the Grand Unified Theories failed. After its successful transition to metaphysics and its self-immunization against empirical control, it is not even clear, if there are any empirical means by which we could find out if string theory does describe nature in any adequate way, or not. Even a wrong theory will not necessarily be defeated by nature, if it is able to avoid any contact to nature.

What, if string theory can stabilize successfully its actual status in the limbo between unattainable empirical confirmation and just as unattainable empirical refutation? Possibly it achieves the stabilization of this status by means of further conceptual metamorphoses.

> *"In spite of the tremendous mental power of the people working in it, in spite of the string revolutions and the excitement and the hype, years go by and the theory isn't delivering physics. All the key problems remain wide open. The connection with reality becomes more and more remote."* (Rovelli (2003) 20)

In such a situation, it seems natural to ask, if physics possibly has taken the wrong turn at a certain point of its development towards a theory of quantum gravity.

> *"Simp - It is not the fault of the theoretical physicist if the path of the natural evolution of the research has lead to a theory which is very complicated.*
> *Sal - And if it was the fault of the theoretical physicist? I suppose when you say 'the path of the natural evolution of the research' you mean the line that goes along Fermi theory, QED, SU(2) x U(1), QCD, the standard model, and then grand unified theories, the revival of Kaluza-Klein, supersymmetry, supergravity, ... strings...*
> *Simp - Yes.*
> *Sal - But what if this 'path of natural evolution' has taken a wrong turn at some point."* (Rovelli (2003) 6)

String theory follows without doubt a strategy drawn out for a long time by the unification program of physics. - On the one hand, as already discussed above, the basic assumption behind the unification program, the idea of an ontic unity of nature, could finally turn out to be wrong here in the context of quantum gravity. The implications of the unification program would turn out to be ultimately inadequate for a description of nature. Or its metaphysical prejudices could, although they might be correct in a certain way, turn out to be only of very limited usefulness for our epistemic endeavours. Quantum gravity would simply be the realm in which the resulting problems become virulent.

On the other hand, the problems could as well be an implication of the mathematical and model-theoretical basis of string theory. String theory uses, as almost all physical theories do, the mathematics of the continuum. There are good reasons to assume that this mathematical apparatus should turn out to be simply inadequate in the context of quantum gravity.[23] Possibly even more problematic is the use of a static background spacetime in string theory.[24] With the insights already achieved in the context of general relativity, this static background spacetime can hardly be part of an adequate theory of quantum gravity. If string theory can be seen in any way as a theory of quantum gravity at all - as an attempt at a (complete) parameterization of its nomological structure or of its spectrum of dynamical expressions -, it is probably a parameterization using a rather inadequate mathematical basis: a parameterization of spatially and structurally discrete phenomena carried out by force within the context of the mathematics of the continuum - the inheritance of quantum field theory, if not of the complete traditional model-building in physics.

The hope remains that the still unknown fundamental principle for string theory, when it finally will be found, could uncover this fact and thereby lead to a more adequate mathematical basis. The necessity of a background independent formulation for quantum gravity, forced by the insights of general relativity, as well as the holographic principle[25] could be of heuristic significance in the search for such a principle.

But it is also possible that such a principle will never be found, because some of the basic assumptions of string theory are simply wrong. Even the idea of a fundamental nomological basis of nature might finally turn out to be wrong in the energy range of quantum gravity, as John Wheeler's *Law-Without-Law* concept[26] or Holger B. Nielsen's *Random Dynamics*[27] suggest. Low-energy structures could, according to Wheeler and Nielsen, be the result of a coarse-graining from a structureless and stochastic chaos on a "fundamental" level. - And there are probably many more alternatives with regard to the physics at the Planck scale which are not compatible with the basic assumptions of string theory (or any other existing approach to quantum gravity).

> *"[...] nature is much more crazy at the Planck scale than even string theorists could have imagined."* ('t Hooft (1993) 2)

---

[23]     Considerations with regard to the Bekenstein-Hawking entropy of black holes lead to the idea that a discrete structure (and finite information densities) should be assumed for the most fundamental level of nature. Cf. Bekenstein (2000) and (2001), Jacobson (1999), Jacobson / Marolf / Rovelli (2005) and Sorkin (2005). In the context of quantum gravity, the mathematics of the continuum could well lead to theoretical artifacts without any descriptive content.

[24]     See above.

[25]     Cf. Bekenstein (2000), Bousso (2002), Horava (1999), Markopoulou / Smolin (1999), Susskind (1995), 't Hooft (1993) and (2000).

[26]     Cf. Wheeler (1979), (1983) and (1989).

[27]     Cf. Nielsen (1983), Frogatt / Nielsen (1991) and Nielsen / Rugh / Surlykke (1994).

*"[...] it is unlikely that a final theory of quantum gravity - if indeed there is one - will look much like any of the current candidate theories, be they string theory, canonical gravity, or other approaches."* (Weinstein (2005) 10)

The already existing alternatives to string theory[28], with the exception of *Loop Quantum Gravity*[29], still do not have such a highly developed and highly complex formal apparatus at their disposal like string theory. And, in contrast to string theory, they can not look back on decades of efforts carried out by many physicists. But this situation of a massive imbalance with regard to efforts already made, time for development, and institutional infrastructure between the different approaches to a theory of quantum gravity could change (at least gradually) in the future, if the fundamental problems of string theory and their implications are taken seriously as motivations for alternative approaches. A scientific monoculture seems hardly adequate at the present situation. String theory has too many severe problems to be seen as the only option for quantum gravity. And to think that the end of physics is coming soon seems completely unreasonable in the present situation.

*"I think we are still very far from the end of physics!"* (Rovelli (2003) 15)

---

[28]    A survey of the spectrum of already existing alternative approaches to quantum gravity can be found in Rovelli (1998).

[29]    Cf. Rovelli (2004).

# References

Ashtekar, A. (2005). Gravity and the Quantum. *New Journal of Physics* **7**, 198; also: arXiv: gr-qc/0410054.

Banks, T. / Dine, M. / Gorbatov, E. (2003). Is there a String Theory Landscape?. arXiv: hep-th/0309170.

Bekenstein, J.D. (2000). Holographic Bound from Second Law. arXiv: gr-qc/0007062.

Bekenstein, J.D. (2001). The Limits of Information. *Studies in the History and Philosophy of Modern Physics* **32**, 511-524; also: arXiv: gr-qc/0009019.

Bousso, R. (2002). The Holographic Principle. *Reviews of Modern Physics* **74**, 825-874; also: arXiv: hep-th/0203101.

Butterfield, J. / Isham, C. (2001). Spacetime and the Philosophical Challenge of Quantum Gravity. in: Callender/Huggett (2001).

Callender, C. / Huggett, N. (Eds.) (2001). *Physics meets Philosophy at the Planck Scale. Contemporary Theories of Quantum Gravity*. Cambridge: Cambridge University Press.

Cartwright, N. (1983). *How the Laws of Physics lie*. Oxford: Clarendon Press.

Cartwright, N. (1989). *Nature's Capacities and Their Measurement*. Oxford: Clarendon Press.

Cartwright, N. (1994). Fundamentalism vs the Patchwork of Laws. *Proceedings of the Aristotelian Society* **93**, 279-292.

Cartwright, N. (1999). *The Dappled World*. Cambridge: Cambridge University Press.

Dawid, R. (2003). Scientific Realism in the Age of String Theory. http://philsci-archive.pitt.edu, Document PITT-PHIL-SCI00001240.

Dienes, K.R. (1997). String Theory and the Path to Unification. A Review of Recent Developments. *Physics Reports* **287**, 447-525; also: arXiv: hep-th/9602045.

Dine, M. (2004). Is there a String Theory Landscape: Some Cautionary Remarks. arXiv: hep-th/0402101.

Douglas, M.R. (2003). The Statistics of String/M Theory Vacua. *Journal for High Energy Physics* **0305**:046; also: arXiv: hep-th/0303194.

Frogatt, C.D. / Nielsen, H.B. (1991). *Origin of Symmetries*. Singapore: World Scientific.

Giddings, S.B. (2005). Gravity and Strings. arXiv: hep-th/0501080.

Green, M.B. / Schwarz, J.H. / Witten, E. (1987). *Superstring Theory*, 2 Vols., Cambridge: Cambridge University Press.

Greene, B. (1999). *The Elegant Universe: Superstrings, Hidden Dimensions, and the Quest for the Ultimate Theory*. New York: W.W. Norton.

Hedrich, R. (1990). *Komplexe und fundamentale Strukturen  -  Grenzen des Reduktionismus*. Mannheim / Wien / Zürich: B.I. Wissenschaftsverlag.

Hedrich, R. (1995). Unsere epistemische Situation, ihre Grenzen und ihre neuronalen Determinanten - Der Objektbegriff. *Philosophia Naturalis* **32/1**, 117-139.

Hedrich, R. (1998). *Erkenntnis und Gehirn - Realität und phänomenale Welten innerhalb einer naturalistisch-synthetischen Erkenntnistheorie*. Paderborn: Schöningh / Mentis.

Hedrich, R. (1998a). Las bases materiales de nuestras capacidades epistémicas - Ensayo de una revisión sintética de la epistemología. *Argumentos de Razón Técnica* **1**, 91-109.

Hedrich, R. (1999). Die materialen Randbedingungen epistemischer Leistungen. *Philosophia Naturalis* **36/2**, 237-262.

Hedrich, R. (2001). The Naturalization of Epistemology and the Neurosciences. *Epistemologia - Rivista Italiana di Filosofia della Scienza / An Italian Journal for the Philosophy of Science* **24/2**, 271-300.

Hedrich, R. (2002). Anforderungen an eine physikalische Fundamentaltheorie. *Zeitschrift für Allgemeine Wissenschaftstheorie / Journal for General Philosophy of Science* **33/1**, 23-60.

Hedrich, R. (2002a). Superstring Theory and Empirical Testability. http://philsci-archive.pitt.edu, Document: PITT-PHIL-SCI00000608.

Hedrich, R. (2002b). Epistemische Grenzen. in: W. Hogrebe (Ed.): *Grenzen und Grenzüberschreitungen - XIX. Deutscher Kongress für Philosophie - Bonn 2002*. Bonn, 317-327.

Hedrich, R. (2005). In welcher Welt leben wir? - Superstrings, Kontingenz und Selektion. in: G. Abel (Ed.): *Kreativität - XX. Deutscher Kongress für Philosophie - Sektionsbeiträge*, Band 1. Berlin, 867-879.

Hedrich, R. (2005a). In welcher Welt leben wir? - Physikalische Vereinheitlichung, Kontingenz und Selektion im Superstring-Ansatz. in: W. Lütterfelds (Ed.): *Vom Sinn und Unsinn des menschlichen Lebens*. Passau, 5-26.

Hedrich, R. (2006). String Theory - From Physics to Metaphysics. arXiv: physics/0604171; also: http://philsci-archive.pitt.edu, Document: PITT-PHIL-SCI00002709; submitted to: *Physics and Philosophy*.

Hedrich, R. (2006a). Kohärenz und Kontingenz - Grundlagen der Superstring-Theorie. in: B. Falkenburg (Ed.): *Philosophie im interdisziplinären Dialog*. Paderborn: Mentis.

Hedrich, R. (2006b). *Von der Physik zur Metaphysik - Physikalische Vereinheitlichung und Stringansatz* (to be published).

Horava, P. (1999). M-Theory as a Holographic Field Theory. *Physical Review* **D 59**, 046004; also: arXiv: hep-th/9712130.

Jacobson, T. (1999). On the Nature of Black Hole Entropy. arXiv: gr-qc/9908031.

Jacobson, T. / Marolf, D. / Rovelli, C. (2005). Black Hole Entropy: inside or out?. *International Journal of Theoretical Physics* **44**, 1807-1837; also: arXiv: hep-th/0501103.

Kaku, M. (1999). *Introduction to Superstrings and M-Theory*, 2nd Ed., New York: Springer.

Lerche, W. (2000). *Recent Developments in String Theory*. Wiesbaden: Westdeutscher Verlag; also: arXiv: hep-th/9710246.

Lüst, D. / Theisen, S. (1989). *Lectures on String Theory*. Lecture Notes in Physics 346. New York: Springer.

Markopoulou, F. / Smolin, L. (1999). Holography in a Quantum Spacetime. arXiv: hep-th/9910146.

Morrison, M. (2000). *Unifying Scientific Theories. Physical Concepts and Mathematical Structures.* Cambridge: Cambridge University Press.

Nielsen, H. (1983). Field Theories without Fundamental Gauge Symmetries. *Philosophical Transactions of the Royal Society London* **A 310**, 261-272.

Nielsen, H.B. / Rugh, S.E. / Surlykke, C. (1994). Seeking Inspiration from the Standard Model in order to go beyond it. arXiv: hep-th/9407012.

Polchinski, J.G. (2000). *String Theory. Vol. 1: An Introduction to the Bosonic String*. Cambridge: Cambridge University Press.

Polchinski, J.G. (2000a). *String Theory. Vol. 2: Superstring Theory and Beyond*. Cambridge: Cambridge University Press.

Rickles, D. (2005). Interpreting Quantum Gravity. http://philsci-archive.pitt.edu, Document: PITT-PHIL-SCI00002407.

Rovelli, C. (1998). Strings, Loops, and the Others: A Critical Survey on the Present Approaches to Quantum Gravity. in: N. Dadhich / J. Narlikar (Eds.): *Gravitation and Relativity: At the Turn of the Millennium*. Poona; also: arXiv: gr-qc/9803024.

Rovelli, C. (2000). Notes for a brief history of quantum gravity. arXiv: gr-qc/0006061.

Rovelli, C. (2001). Quantum Spacetime: What do we know?. in: Callender/Huggett (2001). arXiv: gr-qc/9903045.

Rovelli, C. (2003). A Dialog on Quantum Gravity. *International Journal of Modern Physics* **12**, 1; also: arXiv: hep-th/0310077.

Rovelli, C. (2004). *Quantum Gravity*. Cambridge: Cambridge University Press; also: www.cpt.univ-mrs.fr/~rovelli.

Schroer, B. (2006). String Theory and the Crisis in Particle Physics. arXiv: physics/0603112.

Schwarz, J.H. (Ed.) (1985). *Superstrings: The First 15 Years of Superstring Theory*, 2 Vols., Singapore: World Scientific.

Schwarz, J.H. (1998). Beyond Gauge Theories. arXiv: hep-th/9807195.

Schwarz, J.H. (2000). Introduction to Superstring Theory. arXiv: hep-ex/0008017.

Smolin, L. (2001). *Three Roads to Quantum Gravity*. New York: Basic Books.

Smolin, L. (2003). How far are we from the quantum theory of gravity?. arXiv: hep-th/0303185.

Smolin, L. (2004). Scientific Alternatives to the Anthropic Principle. arXiv: hep-th/0407213.

Smolin, L. (2005). The Case for Background Independence. arXiv: hep-th/0507235.

Sorkin, R.D. (2005). Ten Theses on Black Hole Entropy. *Studies in History and Philosophy of Modern Physics* **36**, 291-301; also: arXiv: hep-th/0504037.

Susskind, L. (1995). The World as a Hologram. *Journal of Mathematical Physics* **36**, 6377-6396; also: arXiv: hep-th/9409089.

Susskind, L. (2003). The Anthropic Landscape of String Theory. arXiv: hep-th/0302219.

Susskind, L. (2004). Supersymmetry Breaking in the Anthropic Landscape. arXiv: hep-th/0405189.

Susskind, L. (2005). *The Cosmic Landscape - String Theory and the Illusion of Intelligent Design*. New York: Little, Brown & Co.

't Hooft, G. (1993). Dimensional Reduction in Quantum Gravity. arXiv: gr-qc/9310026.

't Hooft, G. (2000). The Holographic Principle. arXiv: hep-th/0003004.

Vafa, C. (1997). Lectures on Strings and Dualities, 1996 ICTP Summer School Lectures. arXiv: hep-th/9702201.

Weinberg, S. (1992). *Dreams of a Final Theory*. New York: Pantheon Books.

Weingard, R. (1989). A Philosopher looks at String Theory. *Proceedings of the Philosophy of Science Association 1988*, Vol. 2, Chicago, 95-106; also in: Callender / Huggett (2001).

Weinstein, S. (2005). Quantum Gravity. in: E.N. Zalta (Ed.): *Stanford Encyclopedia of Philosophy*. http://plato.stanford.edu.

Wheeler, J.A. (1979). Frontiers of Time. in: N. Toraldo di Francia (Ed.): *Problems in the Foundations of Physics. Proceedings of the International School of Physics 'Enrico Fermi'*, Course 72. Amsterdam.

Wheeler, J.A. (1983). Law without Law. in: J.A. Wheeler / W.H. Zurek (Eds.): *Quantum Theory and Measurement*. Princeton, N.J.: Princeton University Press.

Wheeler, J.A. (1989). Information, Physics, Quantum: the Search for Links. in: *Proceedings 3rd International Symposium on the Foundation of Quantum Mechanics*. Tokyo, 354-368; also in: W.H. Zurek (Ed.): *Complexity, Entropy and the Physics of Information*. New York: Addison-Wesley (1990), 3-28.

Woit, P. (2001). String Theory - An Evaluation. arXiv: physics/0102051.

Woit, P. (2002). Is String Theory Even Wrong?. *American Scientist* **90**, 110-112.